\documentclass{aa}
\usepackage{psfig, wrapfig}

\begin{document}



   \title{GRB 021219: the first Gamma-Ray Burst localized in real time with IBAS \thanks{Based on observations with INTEGRAL, an ESA project with instruments and science data centre funded by ESA member states (especially
the PI countries: Denmark, France, Germany, Italy, Switzerland, Spain), Czech Republic and Poland, and with the participation of Russia
and the USA. }}

   \author{S. Mereghetti\inst{1}, D. G\"{o}tz\inst{1}$^{,}$\inst{2},
   V. Beckmann\inst{3}$^{,}$\inst{4}, A. von Kienlin\inst{5},
   P. Ubertini\inst{6}, A. Bazzano\inst{6},
   L. Foschini\inst{7}, \and G. Malaguti\inst{7}
}

   \offprints{S. Mereghetti, email: sandro@mi.iasf.cnr.it}

   \institute{Istituto di Astrofisica Spaziale e Fisica Cosmica -- CNR,
              Sezione di Milano ``G.Occhialini'',
          Via Bassini 15, I-20133 Milano, Italy
         \and
             Dipartimento di Fisica, Universit\`{a} degli Studi di Milano Bicocca,
             P.zza della Scienza 3, I-20126 Milano, Italy
         \and
             Integral Science Data Centre, Chemin d'\'{E}cogia 16, CH-1290 Versoix, Switzerland
         \and
            Institut f\"ur Astronomie und Astrophysik, Universit\"at T\"ubingen, Sand 1, D-72076 T\"ubingen, Germany
         \and
             Max Planck Institut f\"{u}r extraterrestrische Physik, Postfach 1312 D-85741, Garching, Germany
         \and
            Istituto di Astrofisica Spaziale e Fisica Cosmica -- CNR, Sezione di Roma, Via Fosso del Cavaliere 100, I-00133 Roma, Italy
         \and
            Istituto di Astrofisica Spaziale e Fisica Cosmica -- CNR, Sezione di Bologna, Via Gobetti 101, I-40129, Bologna, Italy
   }


\abstract{
On December 19, 2002, during the Performance and Verification Phase of INTEGRAL,
a Gamma-Ray Burst (GRB) has been detected and localized in real time
with the INTEGRAL Burst Alert System (IBAS). Here we present the results
obtained with the  IBIS and SPI instruments.
The burst had a time profile with a single peak lasting about 6 s.
The peak spectrum can be described by a single power law with photon
index $\Gamma$=1.6$\pm$0.1 and
flux  $\sim$3.7 photons cm$^{-2}$ s$^{-1}$ (20 - 200 keV).
The fluence in the same energy range  is
9$\times$10$^{-7}$ erg cm$^{-2}$.
Time resolved spectroscopy performed with IBIS/ISGRI shows
a clear hard to soft evolution of the spectrum.
}


\authorrunning{S. Mereghetti et al.}
\titlerunning{INTEGRAL results on GRB 021219}

\maketitle

%

\section{Introduction}
Although considerable progress in the understanding of
Gamma-Ray Bursts (GRBs) has been made after the discovery of their
afterglows at longer wavelengths
(see, e.g., \cite{vpkw})
the debate on their progenitors is still open.
A model involving the core collapse of a massive star (e.g. \cite{woosley})
seems to be favored for long  GRBs  and evidence for a Supernova-GRB association
has been found recently (\cite{stanek,hjorth}).
The situation is more uncertain for what concerns
short ($<$ 2 s) bursts,
mostly because of the lack of counterparts at other wavelengths for this class
of GRBs.

A multi-wavelength approach is crucial to the understanding of the complex nature of GRBs.
The short duration
of the prompt $\gamma$-ray emission and the fading character of the afterglow
impose a rapid follow-up. This
can be achieved only if the positions derived from the prompt emission are
immediately distributed to the scientific community.
The INTEGRAL satellite, although not built as a GRB-oriented mission, can contribute to this
task thanks to the INTEGRAL Burst Alert System (IBAS; \cite{ibas}).
To date IBAS  has successfully detected and located 5 GRBs,
with increasing accuracy and decreasing delays 
(4.36$'$ error radius after 30 s for GRB 030501; IBAS Alert 596).

GRB 021219 is the first GRB detected in real time by IBAS.
This occurred only two  months after the launch of  INTEGRAL,
while the satellite was still in its performance and verification (PV) phase.
During the PV phase, the external distribution
of the IBAS alerts was not enabled yet. An internal alert message,
produced only 10 s after the start of the burst,
reached the members of the IBAS Localization Team and a GRB Coordinates Network
(GCN) circular with a preliminary position could
be issued 5 hours later, after a quick analysis to confirm the  GRB (\cite{mereghetti}).
Since during the PV phase the relative alignment between IBIS and the
satellite star trackers
was not well measured yet,
an uncertainty of 20$'$ was attributed to the derived position.
A refined position could be derived later (\cite{gotz}),
thanks to the presence of  Cyg X-1 in the field of view.

%


Here we report on the results on GRB 021219 obtained with the  IBIS (\cite{ubertini})
and SPI (\cite{vedrenne}) instruments. They are both coded mask imaging telescopes
with a large field of view (29$^{\circ}$$\times$29$^{\circ}$ IBIS,
36$^{\circ}$ diameter SPI).
IBIS is based on two detectors, ISGRI (\cite{lebrun}) and PICsIT (\cite{dicocco}),
operating in the
15 keV - 1 MeV and 175 keV - 10 MeV energy ranges, respectively.
SPI consists of 19 Ge detectors cooled at 85 K and works in the 20 keV - 8 MeV range.
The GRB was located outside
the field of view of the two low energy
monitoring instruments JEM-X (\cite{lund}) and OMC (\cite{mashesse}).


%
\begin{figure}
      \hspace{0cm}\psfig{figure=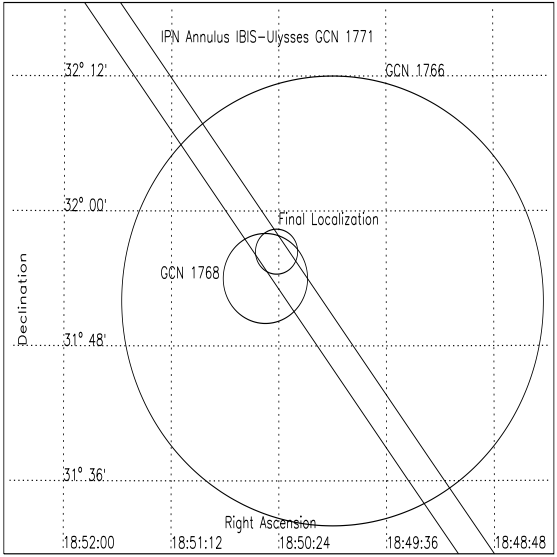,height=8.cm,width=7.5cm}
      \caption[]{Localizations of GRB021219: the positions published earlier are consistent
with the final one derived in this work, see text.}
         \label{ipn}
   \end{figure}

\section{Data Analysis}

\subsection{IBIS}

During this observation PICsIT  provided
images integrated on time intervals of several minutes, which 
cannot be used for the study of GRBs.
Owing to the  telemetry limitations at satellite
level this is the standard operation
mode of PICsIT. 
Useful data for GRB studies can be obtained by PICsIT when it
is operated in photon-by-photon mode (as it occurred for GRB 021125, see \cite{malaguti}).
Our results for GRB 021219 are therefore based only on data from the ISGRI detector.

We have analyzed ISGRI single interaction events,
for which the arrival time, energy deposit and interaction pixel are known.

\begin{figure}
      \hspace{0cm}\psfig{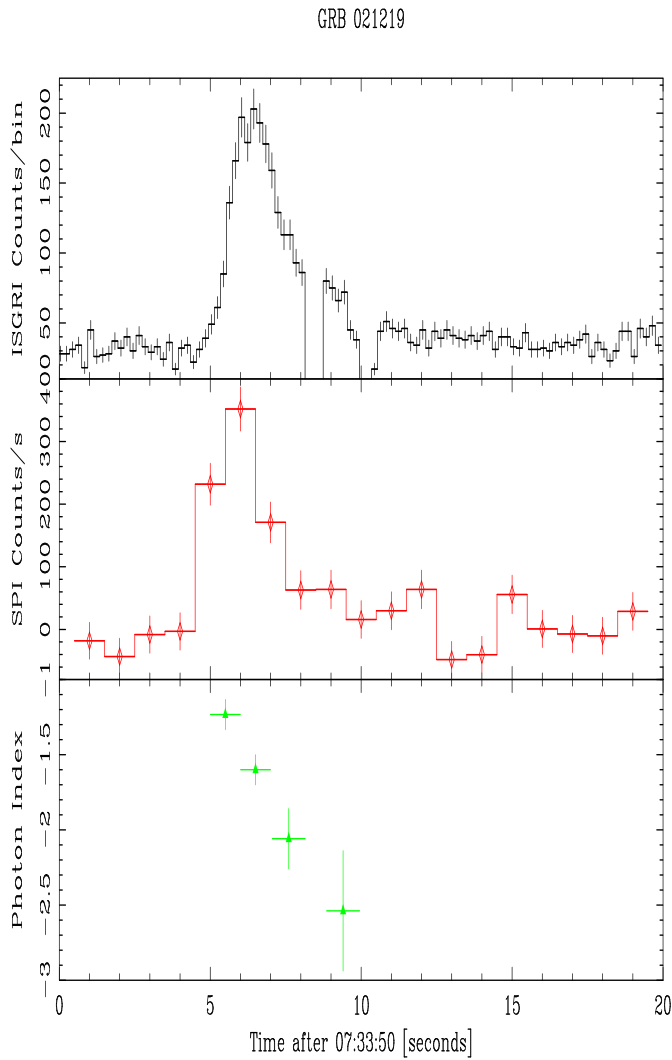}
      \caption[]{Upper Panel: IBIS/ISGRI light curve of GRB021219 in the 15-500 keV band binned over 0.2 seconds. The gaps are artifacs caused by satellite telemetry saturation. Middle panel:Background subtracted light curve of GRB 021219 measured with SPI. Lower Panel: Spectral evolution of GRB021219 with time. A clear hard-to-soft evolution is seen. }
         \label{lc}
   \end{figure}

GRB 021219 has been detected at off-axis angles Z=1.38$^{\circ}$ and Y=10.02$^{\circ}$, in the partially
coded field of view of IBIS, with a signal-to-noise ratio of 15.5 in the 15-100 keV band.
The top panel of Fig. \ref{lc} shows the GRB light curve  binned at 0.2 s intervals.
It refers to the 15-500 keV band, and, to increase the signal to noise ratio,
it has been extracted using only the pixels illuminated by the GRB for at
least half of their surface.
The GRB started at 07:33:55 UTC and had  a T$_{90}$ duration of 5.5 s.
The time profile shows a single peak.

To derive an accurate position  we  selected a 15 s time interval starting at 07:33:55 UTC,
obtaining  $\alpha_{J2000}$ = 18$^h$ 50$^m$ 25$^s$,  $\delta_{J2000}$ = +31$^{\circ}$
56$'$ 23$''$, with an uncertainty of 2$'$ radius.
As shown in Fig. \ref{ipn}, this position is consistent with the one
published earlier and with the   annulus derived independently
with the International Planetary Network (IPN, \cite{hurley}).

\begin{figure}
      \hspace{0cm}\psfig{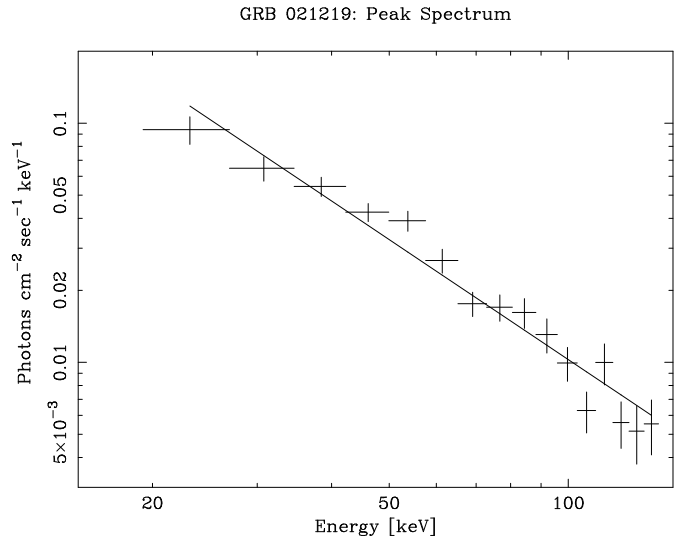}
      \caption[]{Peak spectrum of GRB021219 as measured with IBIS/ISGRI.}
         \label{sp}
   \end{figure}

We have extracted the GRB peak spectrum (over a time interval of 1 s)
using 128 linearly spaced bins between 19 keV and 1 MeV and then rebinned it to have
at least 25 counts per channel.
Since IBIS is a coded mask instrument, the background spectrum can be measured
simultaneously to that of the target, by using the pixels not illuminated by the source.
A fully calibrated response matrix for large off-axis angles is not available yet.
Therefore, we divided the counts spectrum by the closest (in detector coordinates) spectrum of the
Crab Nebula.
The resulting photon spectrum (Fig. \ref{sp}) is well described by a   power law
with photon index $\Gamma$=1.6$\pm$0.1 and flux in the 20-200 keV band
of 3.7 photons (3.5$\times$10$^{-7}$ erg) cm$^{-2}$ s$^{-1}$.

The total (time averaged) spectrum has been obtained with the same technique and
can be fitted by a power law with  $\Gamma$=2.0$\pm$0.1.
The fluence in the 20-200 keV band (corrected for the telemetry gaps)
is 9$\times$10$^{-7}$ erg cm$^{-2}$.

We have also investigated the spectral evolution with time.
Four spectra of the duration of $\sim$1 s each have been extracted: one
during the rising part of the GRB, one at the peak, and two during the decaying tail.
All the spectra are well represented by a power law model.
The corresponding photon indices, plotted in the lowest panel of Fig. \ref{lc},
provide a significant evidence for a softening with time of the burst spectrum.

\subsection{SPI}

At the time of GRB 021219   the SPI instrument was in
low-telemetry mode. Therefore, no single events have been transmitted
to the ground in photon-by-photon mode. Only the events which occurred in
several detectors (multiple
events) and those  analysed by the on-board pulse
discriminator (PSD events) were kept. Due to this, there was no sensitivity
below $\sim 200 \rm \, keV$. Still SPI transmitted to the ground the countrates
of all the events measured in each of the 19 Ge detectors in bins of 1 s
(see middle panel of Fig. \ref{lc}).
For these events the information about the photon
energy is lost and thus only a broad band
count rate can be given. The background was determined using the data starting
from 20 minutes before the burst occurrence.
The peak flux and fluence have been then derived from
the broad band count rate by comparison with the count rate
measured for the Crab Nebula. This results in a peak flux of $10.7 \pm 1.1 \,
\rm Crab$ at 07:33:56 UTC, which
is consistent with the value obtained with IBIS/ISGRI. The values are consistent
also for what concerns the fluence.

The imaging analysis of the SPI data, using a 3 s time interval starting at 07:33:55 UTC,
yields a position $\alpha_{J2000}$ = 18$^{h}$ 49.3$^{m}$,  $\delta_{J2000}$ = +31$^{\circ}$
46.1$'$, with an uncertainty of 30$'$ (S/N = 8.5)
localizing the GRB 17.6$'$ from the IBIS/ISGRI position.

\section{Discussion}

GRB 021219 had a  rather steep average spectrum, indicating that it can be
considered an X-ray rich GRB.
Usually GRB spectra are well described by a phenomenological  model (\cite{band})
consisting of two power laws, one at low energies
(with slope $\alpha$) and one at high energies (with slope $\beta$),
and a  smooth break
typically between about 100 and 400 keV (\cite{preece}).
The fact that no break  is seen in the spectrum of GRB 021219 up to 150 keV indicates
that the break, if any, is outside our detection range. The value of the photon
index $\sim$2, compared to the ones typically observed in the  BATSE sample
$\langle \beta\rangle\sim2.5$,
suggests that we are observing the part of the spectrum above the break.
We cannot exclude, however, that the break is at higher energies: this would
qualify GRB 021219 even more as an X-ray rich GRB.

The time resolved spectroscopy performed with IBIS/ISGRI indicates a
clear hard-to-soft spectral evolution.
This is a common feature in many GRBs observed with previous
satellites (e.g. \cite{norris}, \cite{ford}, \cite{ff}).

No optical counterpart for this GRB has been reported, with limiting
magnitudes of R=13.7 at t-t$_{0}$$\sim$7.5 hours (\cite{lipunov}),
R=18 at t-t$_{0}$$\sim$11 hours,
I=19.5 at t-t$_{0}$$\sim$18 hours (\cite{henden})
and R=20.5 at t-t$_{0}$$\sim$34 hours (\cite{castro}).
Observations in the radio band at 4.86 GHz (at t-t$_{0}$$\sim$16 hours) have not
detected any new source down to a 4$\sigma$ limit of 220 $\mu$Jy (\cite{berger}).

The fact no optical transient  has been detected indicates a "dark'' or at least
dim nature of this GRB.
A similar behaviour  has been noticed before in other X-ray rich events.
For example, the X-ray rich GRB 021211 (\cite{crew})
 was fainter than R$\sim$22.5 12 hours after the burst (\cite{klose}).
The optical time history of this GRB is compatible with the upper
limits reported above for GRB 021219.

\section{Conclusions}

The successful detection and localization of GRB 021219
only one month after IBIS activation,
showed  that IBAS is able to derive and distribute the
position of the GRBs detected in the field of view of IBIS within a few tens of seconds.

The analysis of GRB021219 reported here shows that IBIS/ISGRI is indicated for
detailed spectral studies. In fact we have been able to obtain spectra integrated over only 1 s with good
statistics, thus deriving the spectral shape of the GRB and its
variation with time.

\begin{acknowledgements}
This research has been partially supported by the Italian Space Agency.
LF acknowledges the hospitality of the ISDC during part of this work.
\end{acknowledgements}

\end{document}